%Paper: hep-th/9307053
%From: "Simon Dalley" <dalley@puhep1.Princeton.EDU>
%Date: Wed, 7 Jul 93 21:38:08 EDT

\input phyzzx

\sequentialequations

                        %Real Nucl. Phys. format
\overfullrule=0pt
\catcode`\@=11

\def \F{\phi}

\def \D{\delta}

\def \DM{ {\partial \over {\partial \mu}}}

\def\NP{{\it Nucl. Phys.\ }}

\def\PL{{\it Phys. Lett.\ }}
\def\PR{{\it Phys. Rev.\ }}
\def\PRL{{\it Phys. Rev. Lett.\ }}

\def\IJMP{{\it Int. Jour. Mod. Phys.\ }}
\def\Mod{{\it Mod. Phys. Lett.\ }}

\def\lc{light cone}

\def\eqaligntwo#1{\null\,\vcenter{\openup\jot\m@th
\ialign{\strut\hfil
$\displaystyle{##}$&$\displaystyle{{}##}$&$\displaystyle{{}##}$\hfil
\crcr#1\crcr}}\,}
\catcode`\@=12

\REF\Hoof{G.'t Hooft, \NP {\bf B72} 461 (1974).}
\REF\Wein{D. Weingarten, \PL {\bf 90} (1980) 280.}
\REF\David{F.David, \NP {\bf B257} 45 (1985).}
\REF\Migetc{V.A.Kazakov, I.K.Kostov, and A.A.Migdal,
\PL {\bf B157} (1985) 295.}
\REF\Migkaz{V.A.Kazakov and A.A.Migdal, \NP {\bf B311} (1989) 171.}
\REF\Dall{S. Dalley, \Mod {\bf A7}, 1651 (1992).}
\REF\MD{M.R.Douglas, \PL {\bf B238} (1990) 176.}
\REF\Pen{R.C.Penner, J. Diff. Geom. {\bf 27} (1988) 35.}
\REF\Kont{M.Kontsevich, Commun. Math. Phys. {\bf 147} (1992) 1.}
\REF\DKo{S.Dalley and I.R.Klebanov, \PL {\bf B298} 79 (1993).}
\REF\Brod{H.-C. Pauli and S. Brodsky, \PR {\bf D32} (1985) 1993 and 2001.}
\REF\DKg{S.Dalley and I.R.Klebanov, \PR {\bf D47} 2517 (1993).}
\REF\spoly{A.M.Polyakov, \PL {\bf B103} (1981) 207.}
\REF\div{J.Koplik, A.Neveu, and S.Nussinov, \NP {\bf B123} (1977) 109.}
\REF\CThorn{C. B. Thorn, \PL {\bf 70B} (1977) 85; \PR {\bf D32} (1978) 1073.}
\REF\early{Chang and Ma, \PR {\bf 180} (1969) 1506; J.Kogut and Soper,
\PR {\bf D1} (1970) 2901.}
\REF\kres{K.Demeterfi and I.R.Klebanov, PUPT--1370, hep-th/9301006,
presented at the 7th
Nishinomiya-Yukawa Memorial Symposium ``Quantum Gravity'', November 1992.}
\REF\CT{T. L. Curtright and C. B. Thorn, \PRL {\bf 48}, 1309 (1982).}
\REF\Das{S.Das, A.Dhar, A.Sengupta, and S.Wadia, \Mod {\bf A5} (1990) 1041.}
\REF\migbou{D.V.Boulatov, V.A.Kazakov, I.K.Kostov, and A.A.Migdal,
\NP {\bf B257}
(1985) 641.}
\REF\jan{J.Ambjorn, B.Durhuus, and J.Frohlich, \NP {\bf B257} (1985) 433.}
\REF\EK{T.Eguchi and H.Kawai, \PRL {\bf 48} (1982) 1063.}
\REF\LAG{L.Alvarez-Gaum\'e, C.Crnkovic, and J.F.L.Barb\`on, \NP {\bf B394}
(1993) 383.}
\REF\KS{I. R. Klebanov and L. Susskind, \NP {\bf B309}, 175 (1988).}
\REF\DM{S. Dalley and T.R. Morris, \IJMP {\bf A5}, 3929 (1990).}

\def\eqaligntwo#1{\null\,\vcenter{\openup\jot\m@th
\ialign{\strut\hfil
$\displaystyle{##}$&$\displaystyle{{}##}$&$\displaystyle{{}##}$\hfil
\crcr#1\crcr}}\,}
\catcode`\@=12

\def\a{\alpha}

\def\half{{1\over 2}}
\def\d{\dagger}

\nopagenumbers

{\baselineskip=12pt
\line{\hfil PUPT-1410}
\line{\hfil hep-th/9307053}
%\line{\hfil July 1993}
}
\title{Light-Cone Quantisation of Matrix Models at $c>1$\foot{To appear in
proceedings of the NATO Advanced Workshop on {\em Recent Developements
in Strings, Conformal Models, and Topological Field Theories},
Carg\`ese, France,
12-21 May 1993.}}
\author{Simon Dalley}
\JHL
\abstract
The technique of (discretised) light-cone quantisation, as applied to matrix
models of relativistic strings, is reviewed. The case of the $c=2$
non-critical bosonic string is discussed in some detail to clarify the nature
of the continuum limit. Further applications for the technique are then
outlined.
\endpage

\pagenumbers
\vsize=8.9in
\hsize=6.5in

\centerline{\caps 1. Introduction}
\bigskip

Random surface problems appear in many branches of theoretical and
mathematical physics. A number of them may be reformulated as, or arise
from, matrix field theories [\Hoof, \Wein, \David, \Migetc, \Migkaz, \Dall],
planar diagrams modelling the fluctuating surfaces. The possible physical
applications, together with  relations to the mathematics of systems of
integrable differential equations [\MD] and moduli space of Riemann surfaces
[\Pen, \Kont],
make this a fascinating and novel (ab)use of quantum field theory.
While  condensed matter problems involve the thermal fluctuations of
observable surfaces in 3D Euclidean space, for applications to high-energy
physics the surfaces are worldsheets which enter indirectly
through the weak coupling
expansions of relativistic string theories purporting to describe quantum
gravity and confining gauge theories. Matrix models of relativistic strings
should be defined in Minkowski space, their spectrum, or a self-consistent
truncation of it, comprising string excitations. While knowledge of the
fractal geometry of perturbative diagrams is sometimes useful,
numerical computation of
critical random surface properties in Euclidean space is no substitute for
direct calculation of the relativistic string observables; for example,
the almost uninvestigated issues of strongly coupled string theories would
be otherwise neglected.

I.Klebanov and the author have suggested [\DKo]
that discretised light-cone quantisation (DLCQ), introduced with a view to
calculating the bound-state spectrum of gauge theories directly [\Brod], may
be very appropriately applied to matrix models also. This method, although
primarily a numerical one, paints a clear physical picture of otherwise
complex dynamics. In the case of string theories, this means  the size,
shape, and energy of strings. The unfortunate circumstance that the simplest
bosonic string theory, to which one would add more structure for a
physically realistic object, is (expected to be) tachyonic poses no
particular problem to DLCQ of matrix models. On the contrary, one may
analyse this pathology with some rigour and, as a result, set about curing
it.

The following section contains an elementary review of matrix models,
light-cone quantisation for a $c=2$ model being carried out in section 3.
Section 4 describes the critical behaviour of the discretised version,
while future avenues of development are discussed in section 5.
\endpage

\centerline{\caps 2. Matrix Models of Random Surfaces}
\bigskip

Consider an $N$x$N$ hermitian matrix field ${\F}_{ab}(x)$ in $c$ dimensions
subject to the following (Euclidean) action
$$ S_{E}=\int d^c x \Tr \left (\half (\partial_\alpha \F)^2+\half \mu \F^2-
{1\over 3\sqrt{N}}\lambda \F^3 \right )\ .\eqn\action$$
The Feynman rules are those of ordinary $\F^3$ field theory except that
worldlines are double lines, each line carrying a ``colour'' index $a$.
Colour is conserved along the propagator and at vertices on account of the
global $U(N)$ symmetry, $\F \to \Omega^{\dagger} \F \Omega$, of the
action;
$$ <\F_{ab}(x) \F_{cd}(y) > = \D_{ad} \D_{bc} \int {d^c p \over (2\pi)^c }
{{\rm e}^{i p \cdot ( x - y) } \over p^2+\mu}\ ,\eqn\prop$$
$$ < \F_{ab}(x) \F_{cd}(x) \F_{ef}(x) > = \D_{bc} \D_{de} \D_{fa} N\
.\eqn\vert$$
The $1/N$ expansion of the theory described by \action\ is a topological
expansion in surfaces [\Hoof], the Feynman diagrams being understood as
fishnet approximations drawn on continuous 2-dimensional surfaces embedded
in $R^c$;
$$ \int {\cal D} \F {\rm e}^{-S_{E}} \sim
\sum \lambda^{v} \left( {1 \over N} \right)^{-
\chi } \int {\rm embeddings}\ .\eqn\form$$
$v$ is the number of vertices, $\chi$ the Euler number, and one integrates over
all embeddings of the graph with rules \prop\ \vert\ . The discretised surface
picture is formalised by considering the dual graphs [\David, \Migetc]
that join centres of
neighbouring loops, which for a $\F^3$ theory specifies a simplicial
triangulation. Each triangle carries unit intrinsic area and its centre
coincides with the Feynman vertex; in particular this leaves the angle
between neighbouring triangles unspecified in this model. The idea is then to
tune the coupling $\lambda$ to a critical value $\lambda_{c}$ at which surfaces
of large intrinsic area $v$ are favourable. At this point one may be able to
take a continuum limit for surfaces.\foot{There may be no scaling of the
intrinsic geometry (curvatures, etc) even though the area scales -- the
$c\leq 1$ models are examples of this -- while at $c>1$ there is even a
danger, commonly attributed to tachyons, of non-scaling area.}

The observables are given by the Green's functions of the matrix field
theory. A
closed string {\em state} is a hole cut into the surfaces at some fixed time,
$t_{0}$ say. Such states are given by the singlet operators in the matrix
model;
$$ {\rm Tr} [ \F (x_{1}) \ldots \F (x_{B}) ]_{t=t_{0}} \eqn\state$$
is a $B$-bit string, the bits being embedded at $x_{1}, \ldots , x_{B}$
respectively and forming a closed chain dual to the external legs of the
Feynman diagrams. Also present are direct products of \state\
(multi-string states) and non-singlets; the latter do not have an
interpretation as closed strings and should be eliminated, either by performing
a self-consistent truncation or effecting a dynamical decoupling (by gauging
the $U(N)$ for example [\DKg]).
The critical point $\lambda_{c}$ of large surfaces
should manifest itself as singularities of Green's functions, such as the
string propagator $\sim < {\rm Tr} [\F \cdots \F ] {\rm Tr} [\F \cdots \F ]
>$, and thus in particular the spectrum of string excitations \state\ should
exhibit some sort of critical behaviour. Applying DLCQ to the matrix field
theory in Minkowski space, one derives the spectrum as a function of $\lambda$
and can search for such behaviour.

The simplest non-trivial example to consider is $c=2$, for which \action\
models, in its $1/N$ expansion, a discrete version of the worldsheet action
for the $c=2$ non-critical bosonic string [\spoly],
$$A = \int d^2 \sigma \sqrt{-\det{g}} (\Lambda + \nu R(\sigma )
+ T \sum_{\mu,\nu = 1}^{2} \eta_{\mu \nu} g^{\alpha \beta}
\partial_{\alpha} x^{\mu} \partial_{\beta} x^{\nu} + O[(\partial x)^4])\
.\eqn\wsheet$$
$\Lambda, \nu$, and $T$ are renormalised parameters associated with $\lambda,
1/N$, and $\mu$ respectively. Stretching energy of the worldsheet is governed
by the propagator \prop\ , which specifies the probability amplitude for the
separation of centres of neighbouring triangles. This exponential fall-off
gives the Gaussian term in $A$ plus non-renormalisable higher
derivative terms. Naively the latter are irrelevant, but such reasoning
assumes scaling of the worldsheet in some sense. The $c=2$ theory \action\
is rendered perturbatively finite  by normal ordering, this being
necessary in any case to eliminate graphs dual to pathological triangulations
where two or more sides of the same triangle are identified. According to
general arguments [\div], the sum of graphs at a given order in $1/N$ in a UV
finite theory is also finite for sufficiently small coupling constant; the
$1/N$ expansion itself is only asymptotic. As $\lambda \to \lambda_{c}$ one
then approaches the edge of the domain of convergence. The phenomenon is
similar to, but certainly different from, the crossover to non-borel
summability in the non-matrix field theory ($N=1$).
\bigskip

\centerline{\caps 3. Light-Cone Quantisation}
\bigskip

Let us now rotate $x^{0} \to ix^{0}$ and find the relativistic string
spectrum by light-cone quantisation. Defining light-cone variables
$x^{\pm}=(x^0\pm x^1)/\sqrt 2$, with $x^{+} = x_{-}$, the light-cone energy
and momentum $P^{\pm} = \int dx^{-} T^{+\pm}$ are
$$ \eqalign{&P^{+}(x^{+}) = \int dx^{-} \Tr (\partial_{-}\F )^{2}\ ,\cr
&P^{-}(x^{+}) = \int dx^{-} \Tr (\half \mu \F^2 - {\lambda \over 3\sqrt{N}}
\F^3 )\ .\cr }\eqn\pminus$$
The light-cone Hamiltonian $P^{-}$ propagates a field configuration from
one $x^{+}$ slice to another. Choosing a free field representation at
$x^{+}=0$ say\foot{The symbol $\dagger$ is always understood
to have purely quantum meaning and never acts on indices.},
$$\F_{ij}={1 \over \sqrt{2\pi}} \int_{0}^{\infty} {dk^{+} \over \sqrt{2k^{+}}}
(a_{ij}(k^{+}){\rm e}^{-ik^{+}x^{-}} + a_{ji}^{\d}(k^{+})
{\rm e}^{ik^{+}x^{-}})\ ,\eqn\Four$$
and imposing the canonical commmutation relations at equal $x^{+}$ lines,
$$ [\F_{ij}(x^{-}),\partial_{-}\F_{kl}(\tilde{x}^{-})] = {i\over 2}
\delta (x^{-} -
\tilde{x}^{-})\delta_{il} \delta_{jk}\ ,\eqn\rel$$
the modes $a_{ij}$ satisfy standard commutators;
$$[a_{ij}(k^{+}),a_{lk}^{\d}(\tilde{k}^{+})] = \delta(k^{+} - \tilde{k}^{+})
\delta_{il}\delta_{jk}\ .\eqn\modeccr$$
An important feature is that longitudinal momentum $k^{+}$ is positive
semi-definite for positive energy quanta\foot{This positivity constraint
results in the extra $1/2$ in \rel\ , ensuring for example that $[P^{+},
\F ] = \partial_{-} \F$.}
In the Fock space constructed using \modeccr\ , the orthonormal single
closed-string states are the singlets
$$N^{-B/2} \Tr [a^{\d}(k_{1}^{+})\cdots a^{\d}(k_{B}^{+})] |0>\ ,\
\sum_{i=1}^{B} k_{i}^{+} = P^+\ .\eqn\string$$
If we let $N \to \infty$ the multi-string states can be neglected, since
$1/N$ is the string coupling constant, and $P^-$ propagates without splitting
or joining strings. Thus we solve for the free string spectrum. (Non-singlets
states are discarded by hand.) $P^+$ and $P^-$ can be simultaneously
diagonalised and so  eigenfunctions of $P^-$ with given $P^+$
will be some superposition
$$\Psi = \sum_{B} \int_{0}^{P^+} dk_{1} \cdots dk_{B} \delta (P^{+} -
\sum k_i ) f_{B} (k_{1}, \ldots ,k_{B}) N^{-B/2} {\rm Tr}[a^{\dagger}(k_{1})
\cdots a^{\dagger} (k_{B}) |0>\ .\eqn\super$$
Explicitly one finds
$$\eqalign{
&:P^{-}: =
\half \mu \int_{0}^{\infty} {dk^{+}\over k^{+}} a_{ij}^{\d}(k^{+})
a_{ij}(k^{+}) -{\lambda \over 4\sqrt{N\pi}}\cr
&\times\int_{0}^{\infty}
{dk_{1}^{+}dk_{2}^{+}\over \sqrt{k_{1}^{+}k_{2}^{+}(k_{1}^{+} + k_{2}^{+})}}
\left\{a_{ij}^{\d}(k_1^+ +k_2^+)a_{ik}(k_2^+)a_{kj}(k_1^+) +
a_{ik}^{\d}(k_1^+)a_{kj}^{\d}(k_2^+)a_{ij}(k_1^+ +k_{2}^+)\right\}
\cr }\ ,\eqn\nominus$$
where repeated indices are summed over.
For the free string theory each of the terms in \nominus\ has a simple local
action on the string. The mass term acts as a tensional energy $\sim \mu
\sum_{i=1}^{B} 1/k_{i}$ and does not change the state. The cubic term can
coalesce two neighbouring bits in the trace, or perform the inverse process,
and changes both the length and momentum distribution (structure function)
of the string. Since $c=2$, there are no transverse oscillations in the
target space. Unlike $c\leq 1$ however, we are dealing with truely stringy
degrees of freedom -- an infinite number of particle fields --
which supplement the centre of mass motion.

Free string states will satisfy a relativistic dispersion relation $2P^+ P^-
=M^2$ and by diagonalising $P^-$ in the basis of states of a fixed $P^+$
we can find the spectrum of masses $M^2$. This diagonalisation cannot be
performed analytically for the full Hamiltonian in general, so we must
introduce a cutoff [\Brod, \CThorn],
rendering the number of states of momentum $P^+$ finite,
and compute numerically. It is therefore important that we can neglect zero
momentum modes $a^\dagger (0)$ since we could in principle include
arbitrary numbers of them without changing $P^+$. Such modes have infinite
energy for $\mu \neq 0$, according to \nominus\ ; however, they can be
infinite in number and so a more careful analysis is required to lift this
ambiguity. Indeed, it is believed by some that a proper constrained
quantisation of these
zero modes is necessary to describe spontaneous symmetry breaking and other
non-perturbative effects in light-cone formalism. In those cases it is
argued that the true
vacuum, if one exists at all, is not the Fock one, $:P^- : |0> =0$,
but a condensate of $a^\dagger
(0)$ modes. But one
should recall that we are interested in \action\ only insofar as it
generates random surfaces through the $1/N$ expansion and perturbation
theory. Therefore we should always take the $\F = 0$ vacuum, for which
there is no condensate of zero modes, since this is the one with respect
to which the planar diagrams are defined. For the $c=2$ model, at each
order in $1/N$ we work with a convergent perturbation expansion, the
non-perturbative effects presumably manifesting themselves through the
${\rm e}^{-N}$ corrections to the asymptotic expansion in $1/N$.
The latter are non-perturbative effects of
string theory and specifying their details is equivalent to stating how
one is going to stabilise the unbounded ${\F}^3$ theory (without
disturbing the $1/N$ expansion), a question that will not be considered
here.

Another useful way of viewing  these and other issues  is to consider
triangulations
in light-cone perturbation theory. Indeed, for relativistic strings
 one could have set up the random surface expansion from this vantage
point from the very beginning.
Using Feynman's causality prescription on $x^+$ rather than $x^0$, the
single-particle propagator becomes
$$\lim_{\epsilon \to 0} \left( \int_{\epsilon}^{\infty}
 {dp^+ \over
4\pi ip^+ }
 \theta (x^+ ) {\rm e}^{-i(x^- p^+ + x^+ \mu/2p^+ )}
+ \int_{-\infty}^{-\epsilon} {dp^+ \over
4\pi ip^+ }
  \theta (-x^+ ) {\rm e}^{-i(x^- p^+ - x^+ \mu/2p^+ )} \right)
+{\delta (x^+ ) \over 2\pi \mu}\ .\eqn\litep$$
The first two terms can be given the usual particle and anti-particle
interpretation by viewing the negative energy $(p^- )$ states as propagating
backwards in time $(x^+ )$. The third term is a special contribution from
$p^+ =0$ and corresponds in the dual diagram to the propagation of a zero
momentum string-bit; neglecting this case,
all particles and anti-particles move
forward in $x^+$ carrying positive $p^+$. In early work [\early], the
quantisation procedure used was the one adopted here, where it was proved
not only that $x^{+}$-ordered and $x^{0}$-ordered perturbation theory are
equivalent, but also that the third term eventually does not contribute
in non-vacuum diagrams. This seems to indicate once again that, provided
we restrict to perturbation theory of \action\ ,
the zero modes can be ignored in
computing the string propagator.
\bigskip

\centerline{\caps 4. Discretisation and Critical Behaviour}
\bigskip

The desired cut-off will be introduced by compactifying $x^-$ and imposing
periodic boundary conditions, $M_{ij}(x^-)=M_{ij}(x^- +L)$ [\Brod]. Then the
 allowed momenta are labelled by positive integers $n_i$;
$$ k^{+}_{i} = {2\pi n_i \over L}\ ,\ P^+ = {2\pi K \over L}\ ,\
\sum_{i=1}^{B} n_{i} = K\ .\eqn\stuff$$
For fixed $P^+$, removing the cut-off $L \to \infty$ corresponds to sending
$K \to \infty$. The ``harmonic resolution'' $K$ represents the total number
of momentum units available to the string. The longest string has $K$ bits
of one unit\foot{This sector alone represents what one might call ``critical
string theory'' [\KS]}, the shortest one bit of $K$ units,
and in general the states can be labelled by
the ordered partitions of $K$ modulo cyclic permutations.
Light-cone quantisation \Four\ - \nominus\ may now be repeated for discrete
variables $k^{+} \to n$ and one finds the mass relation
$${2P^{+}P^{-} \over \mu} = K (V -x T) \ ;\ x={\lambda \over 2\mu \sqrt{\pi}}
\ .\eqn\ham$$
$V$ is the discrete version of the mass term in \nominus\
while $T$ is the cubic term. For finite $K$ the r.h.s. is a finite-dimensional
symmetric matrix with real dimensionless elements -- $\mu$ is the quantity
of dimension ${\rm mass}^2$ which plays the role of string tension
$1/{\alpha}'$ -- which may be diagonalised as a function of the dimensionless
parameter $x$; $V$ is
diagonal while $T$ is off-diagonal.

The following picture of the critical behaviour, supported by the numerical
results [\DKo, \kres], begins to emerge.
In the \lc\ formalism of critical string theory the longitudinal momentum
supported between two points on the string is proportional to the amount of
$\sigma$-coordinate space between these points.
We can adopt a similar co-ordinate
system for the non-critical strings \state\ . Indeed, fixing a particular
bit as origin, we can define a positive scalar field on this $\sigma$-space
by $X=\Delta b/\Delta \sigma$, where $b$ is the  distance,
measured in number of bits, from the origin.
For example, the zero mode $\int X d\sigma$ is the intrinsic length of the
string.
As we remove
the cutoff on the longitudinal momentum allowed for bits ($K \to \infty$),
and hence on
discreteness of $\sigma$-space,
the scalar field will generically take constant values almost
everywhere in $\sigma$-space. We would like to be able to tune the theory
to a critical point where the scalar field is in a long wavelength regime.
In this case it would be somewhat  similar to the Liouville
mode of Poyakov's string [\spoly].
What would this long wavelength regime mean for the spectrum? Firstly
we would expect long string dominance; the expectation value of length for
low-lying  eigenstates would typically diverge as $x \to  x_{c}$ (and we
have been
assuming that this is not distinct from $\lambda_{c}$ discussed earlier).
Since,
roughly speaking, each string-bit carries finite energy, we expect that
$|M^{2}| \to \infty$ as a result. $M^2 \to \infty$ at $x=0$ for infinitely
long strings, but as $x \to x_{c}$, if only for consistency, we must see
$M^2 \to -\infty$ for the low-lying eigenstates if they are long. Indeed
this will tend to happen to the lowest eigenvalues of any real symmetric
matrix as one increases its dimension for sufficiently large off-diagonal
elements. Only if the Hamiltonian is an explicitly bounded operator
combination
$(H \sim O^{\dagger} O + {\rm const})$ can this instability be avoided in
general. We might also expect to see a continuous spectrum at $x=x_{c}$
if we compare with the Liouville theory results [\CT]
$$2P^+P^-\alpha' = p_{\F}^{2}+4 r-{1 \over 6} \ ,
\eqn\Char$$
but it has been difficult to confirm this numerically. Moreover the
groundstate has finite negative mass squared in \Char\ , while the matrix
model's is infinitely negative at $x=x_{c}$. It has been suggested [\kres] that
perhaps $\mu$ should be renormalised to zero as a result, but a derivation
of this requirement is still lacking. In any case we have the first
direct demonstration that the non-critical bosonic string with area
action is tachyonic above $c=1$. The use of an unbounded ${\F}^3$ matrix
potential does not {\em a priori} spell tachyons at any order in $1/N$
in the expansion about $\F=0$, but the string theory described at the
critical coupling is nevertheless tachyonic.\foot{The use of an unbounded
potential is rather a symptom of the divergence of the $1/N$ string
perturbation expansion, as commented earlier.}
\bigskip

\centerline{\caps 5. Future Directions}
\bigskip

In order to identify the critical behaviour at $c=2$ more precisely it is
useful to investigate the effects of adding an explicit polymerisation term
[\Das]
$$S = S_{E} + \int d^2 x\ {g \over N^2} ({\rm Tr}[{\F}^2])^2\ .\eqn\poly$$
For sufficiently large $g$ this worldsheet contact interaction seems to favour
short strings. This is quite unlike the critical point at $g=0$ and, assuming
that there is a phase transition somewhere in between, casts doubt on branched
polymer behaviour of the $c=2$ matrix model in Minkowski space;
recall that this behaviour was
identified in $c>1$ Euclidean dimensions from
numerical simulations [\migbou] and combinatorial estimates
[\jan] of dynamical triangulations at the critical point. Moreover simple
polymerisation is not the only possibility in the Euclidean game.
The phase diagram
needs to be investigated in more detail before a clearer picture can be
gained.

The tachyon we found is expected to persist at $c>2$. To regulate the
transverse dimensions we can use a transverse lattice. To
eliminate the zero modes associated with $x_{\perp}$, which would otherwise
obscure the stringy part of the spectrum, we can perform the Eguchi-Kawai (EK)
compactification [\EK] to single links in each direction. The resulting
field theories are much the same as the $c=2$ one; namely, we deal with
UV finite two-dimensional field theories with convergent perturbation
expansions at given order in $1/N$. If we use Hermitian matrix models
\action\ the EK reduction  induces more general interactions $V_{\rm eff}
(\F )$ in the effective $c=2$ potential [\LAG].
Unfortunately $V_{\rm eff} (\F )$
is not known explicitly and
can only be calculated as an expansion in powers of the inverse transverse
lattice spacing $1/a$. Truncating the expansion arbitrarily at some order,
the resulting model will exhibit  polymeric behaviour for
sufficiently small $a$ since the leading term is a contact interaction
similar to that appended in \poly\ , the induced $g$ being $\sim 1/a^2$.
At sufficiently large $a$ the (unreduced) transverse lattice sites are
obviously uncoupled and we should recover copies of the $c=2$ model at
each site.

In order to deal with an exact EK reduced system, one can employ complex
matrix models similar to the old Weingarten model [\Wein]. Complex matrices $M$
live
on the links of a $c$-dimensional hypercubic lattice. For non-critical
string theory we must use an action given by traces around all
(oriented) loops
of length 4 on the lattice [\Dall],
which comprises the standard plaquette action
plus zero area loops. Expansion of this theory in the coupling constant
and $1/N$ reproduces
the random surfaces of a dynamical quadrangulation in $c$ dimensions, the
the probability distribution for neighbouring vertices $\{x,y\}$
of the quadrangulation
 in this case being
$$\sum_{\hat{\mu}} \delta (x-y -a\hat{\mu})
+ \delta (x-y+a\hat{\mu})\
,\eqn\latprop$$
for orthonormal vectors $\bf{\hat{\mu}}$.
For $c=1$ it has been proven that this gives
the same answers at the critical point as using Feynman propagators
[\Dall].
To perform DLCQ
we must take the naive continuum limit for two of the dimensions, which
seems to produce an intractable two-dimensional kinetic term unfortunately.
Instead we could  use the Feynman propagator for these two continuous
dimensions and study actions like [\KS, \DM]
$$\int d^2 x {\rm Tr}[ \partial_{\a}M \partial^{\a} M^{\dagger} + \mu M
M^{\dagger} + \lambda_{1} MMM^{\dagger}M^{\dagger} + \lambda_{2} MM^{\dagger}M
M^{\dagger}]\ .\eqn\suskle$$
This is a $c=3$ model with EK reduction of the transverse direction
 -- a single complex matrix $M$ lies on
the periodic link  -- particularly simple since
it has no standard plaquette term. One may now study the  DLCQ as a function
of $a$. At sufficiently small $a$ there should be a roughening transition
from the $c=2$ to the $c=3$ phase. It should be noted however that the
computational accuracy diminishes significantly for $c>2$ due to the increase
in degrees of freedom,  each string bit now carrying at least one more
$Z_{2}$ variable -- e.g. the real and imaginary parts of $M$.

Even if reliable data on models such as \suskle\ can be collected, we still
expect tachyons. We must try other possibilities to eliminate them, such as
introducing some dependence on extrinsic geometry, which is also of
interest in condensed matter problems. This is certainly possible
for the
complex matrix models, at least on the lattice target space, since the
orientation of simplices is rather manifest. At a more fundamental level,
we should
look for tachyon-free matrix models with long string dominance if we wish
to describe interacting continuum strings. As indicated earlier, such would
be a delicate theory; using a positive definite Hamiltonian tends
to counteract precisely the requirement of the critical point -- crossover
to divergence of perturbation theory. Also the obvious fact should be
stressed that, contrary to the popular trends of research,
such a theory need not possess
an attendant  continuum worldsheet expansion to be a sensible string theory.

 For other applications of light-cone matrix
models, such as to confining gauge theories, continuity of strings
may not be so
important. For example the interesting Regge-like trajectories found in
two-dimensional gauged matrix models [\DKg] are most probably the result
of restriction to
sectors of fixed (discrete) string length\foot{In ref.[\DKg] low-lying mass
eigenstates tended to consist of strings of some given length for
mysterious dynamical reasons.}.
According to the suggestion made earlier, this
freezes the Liouville-like zero mode and exposes the quasi-harmonic  energy
levels. While it is unrealistic to expect solution of the gauged models in
higher dimensions, since they are more complicated than pure large-$N$ QCD,
the two-dimensional models may help us to understand more clearly the
relationship between gluonic and fundamental strings. They describe a limit
of higher dimensional gauge theory in which all transverse directions are
very compact; $x_{\perp}$-independent
transverse potentials $A_{\perp}(x^+,x^-)$ play
the role of matter
$\F$ in the gauged $c=2$ matrix model.

Clearly there are many interesting questions in applications of
string theory which may be
addressed by the  light-cone matrix models  through analytic and numerical
techniques.

\ack
It is a pleasure to thank  I.Klebanov for many interesting discussions.
Financial support comes through S.E.R.C.(U.K.) post-doctoral fellowship
RFO/B/91/9033.
\endpage
\refout

\bye